\documentclass[12pt,a4paper]{article}
\usepackage[utf8]{inputenc}
\usepackage[spanish]{babel}
\usepackage{amsmath}
\usepackage{amsfonts}
\usepackage{amssymb}
\usepackage{makeidx}
\usepackage{graphicx}
\usepackage{lmodern}
\usepackage{undertilde}

\usepackage[left=2cm,right=2cm,top=2cm,bottom=2cm]{geometry}
\author{ Ricardo Rosas-Rodríguez and Iván Cortes-Cruz\\
\begin{small}
Universidad Tecnológica de la Mixteca
 \end{small}\\ 
\begin{small}
 Instituto de Física y Matemáticas
\end{small}
  }
\title{Cosmological functions and their relation.}
\begin{document}
\maketitle
\begin{small}
{\bf Abstract.} In the non-metric gravity proposed by K. Krasnov, this is a theory in which the scalar constraint of the Ashtekar´s formalism is modified in such way that the cosmological constant is replaced by a cosmological function that depends of the canonical variables $A_{a}{}^{i}$ and $E_{i}{}^{a}$, the General Relativity (GR) is the particular case when the cosmological function is a constant. Some years ago inspired by this theory Rosas-Rodriguez proposed two cosmological functions, one for the Ashtekar´s formalism and the other for the ADM formalism. In this paper we show that this cosmological functions are related thought the three-dimensional Ricci scalar.
\end{small}

\section{Introduction}
It is well know that today there are many problems to face off in General Relativity (GR) \cite{wald, baez, gambi}, for example, we don't know why the observed cosmological constant is so different from the one that is expected to arise as the vacuum energy of quantum fields. For this (and others) reason modified gravity theories have become popular recently \cite{dicke, mann}. In 2008, K. Krosnov introduced a new class of modified gravity theories with the name of non-metric gravity \cite{kr3}. The canonical description of this class of theories shows that as in the usual GR-Pleba\'nsky-Asthekar case the phase space is parametrized by the canonically conjugate pairs $(A_{a}^{i}, \tilde{E}^{ai})$, where $A_{a}^{i}$ is an {\bf su(2)}-connection, $\tilde{E}^{ai}$ is its canonically momentum and the theory is fully constrained \cite{kr2}. Thus at canonical level the only modification with respect to the usual GR is that the cosmological constant gets replaced by a non-trivial (and arbitrary) function $\phi: T^{*}(\mathcal{A}) \rightarrow \mathbb{R}$, where $\mathcal{A}$ denotes the space of all the {\bf su(2)}-connections, of the canonical variables. Inspired by this new class of theories Rosas-Rodríguez around of 2014 proposed a cosmological function for the Ashtekar's formalism that seems to simplify the way what we look for states that solve the constrictions of the theory \cite{rosas1}. A few years ago was proposed a new cosmological function for the ADM formalism which turns out to have the same role that the function for the Ashtekar's formalism \cite{rosas2}. What we do here is give a relation between the proposed cosmological functions. As we will see these cosmological functions are related through the spacial Ricci scalar.

\section{Non-Metric Gravity}
It is well known that Ashtekar's constrain algebra arises as a result of the $3+1$ decomposition of the Pleba\'nski formulation \cite{pleva} of GR. Here we describe the spacetime covariant theory that leads to a kind of modified constraint algebra with respect to the algebra of the Ashtekar's formalism. This theory was proposed in \cite{kr1} and its action has the form
\begin{equation} \label{k12}
S= \int_{\Sigma} B_{i} \wedge F^{i} ( A ) + \frac{1}{2} ( \Psi_{ij} - \frac{1}{3} \delta_{ij} \phi ( \Psi )) B^{i}\wedge B^{j},
\end{equation}
where $ F^{i}$ is the curvature associate to the {\bf su(2)}-connection $A^{i}$, $B^{i}$ is a $2$-form with values in  {\bf su(2)} and $ \Psi^{ij}$ is the Lagrange multiplier field, which is required to be symmetric and traceless. \\

The $3+1$ decomposition proceeds as in \cite{capo} and is given explicit in \cite{kr2} by
\begin{equation} \label{k13}
S= \int_{\Sigma} \int_{\mathbb{R}} d^{3}x dt [ \tilde{E}_{i}^{a} \dot{A}_{a}^{i} + A_{0}^{i} \mathcal{D}_{a}\tilde{E}_{i}^{a} + (\delta^{ij} \utilde{N} + \frac{1}{2} \epsilon^{ijk} \utilde{N}_{k})\tilde{\Psi}_{ij} - \utilde{N} \phi(\Psi) \text{det}\tilde{E}^{ai} ],
\end{equation}
where $\tilde{\Psi}^{ij}:= F_{ab}^{i} \epsilon^{j kl} \tilde{E}_{k}^{a} \tilde{E}_{l}^{b}$, $\utilde{N}^{k}$ is the shift vector, $\utilde{N}$ is the lapse function, and $\tilde{E}^{ai}$ is the momentum conjugate to $A_{ai}$.\\
Thus, the constraints of the theory can be written as
\begin{equation} \label{k2}
\mathcal{G}_{i}=\mathcal{D}_{a} \tilde{E}_{i}^{a}\approx 0 , 
\end{equation}
\begin{equation}\label{k3}
\mathcal{V}_{a}= \tilde{E}_{i}^{b} F_{ab}^{i} \approx 0 , 
\end{equation}
\begin{equation}  \label{k4}
\mathcal{H}= \frac{1}{2}\epsilon^{ijk} \tilde{E}_{i}^{a} \tilde{E}_{j}^{b} F_{ab \, k} - \phi\left( \Psi \right) \text{det}\tilde{E}^{ai} \approx 0 .
\end{equation}
These constraints still form a set of first class, an explicit proof is given in \cite{kr2, rosas3}, which means that the count of the number of physical degrees of freedom (DOF) is unchanged and the theory still propagates two DOF.
Note that the cosmological function $\phi$ depends of the (symmetric) tensor 
\begin{equation} \label{k1}
\Psi^{ij}:=  \frac{1}{\text{det} \tilde{E}^{ai}} \left(  F_{ab}^{i} \epsilon^{jkl}  \tilde{E}_{k}^{a} \tilde{E}_{l}^{b} \right)_{tr-free} , 
\end{equation}
where \textit{tr-free} denotes the trace-free part.\\

Now, we should expect to get also a kind of modified constraint algebra with respect to the constraint algebra of the ADM formalism extended to triads, after all the ADM and Ashtekar´s formalism are related through a canonical transformation. This means that we should have constraints of the form
\begin{equation} \label{cp1}
L^{i} := \epsilon^{ijk} p_{j}^{a} e_{ak} \approx 0,
\end{equation}
\begin{equation} \label{c3}
\mathcal{H}_{a} := - e_{ai} D_{b} p^{bi} \approx 0,
\end{equation}
\begin{equation} \label{c2}
\mathcal{H} : =  p^{ij} p_{ij}  - \frac{1}{2} (p_{i}^{i})^{2} -  4^{\, \, (3)}e^{2 \, \, \, (3)} R  - 8 ^{(3)}e^{2}, \Lambda(\Psi) \approx 0,
\end{equation}
where $\Lambda(\Psi)$ now denote the cosmological function.\\

Inspired by this class of modified theories, Rosas-Rodríguez proposed a few years ago an couple of cosmological functions that appears to avoid some of the problems related to solving the constraints in GR at classical and quantum levels. This cosmological are the subject of the following section.

\section{Cosmological functions}
Note that the cosmological function given by Krasnov $\phi(\Psi)$ are generic, there is not a specific form for these today (however, see \cite{capomon} ). But these cosmological functions gave the idea to Rosas-Rodriguez \cite{rosas1} to suggest the following ansatz as cosmological function in the Astekar's formalism
\begin{equation} \label{k14}
\varphi : =  \frac{1}{2\, \text{det} \tilde{E}^{ai}}  \epsilon^{ijk} \tilde{E}_{i}{}^{a} \tilde{E}_{j}{}^{b} F_{ab}{}_{k}.
\end{equation}
Note that is a scalar function. With this function the Hamiltonian constraint is satisfies automatically at classical and quantum levels. Note that this cosmological function is not a member of the Krasnov's theories. \\
Thus, the problem of find physical states $\Psi[A] \in \mathcal{H}_{phy}$, where $\mathcal{H}_{phy}$ denotes the physical states space, is reduced. There is no problem with the Hamiltonian constraint since any state that satisfies the Gauss and vectorial constraints, 
\begin{equation} \label{k15}
\widehat{\mathcal{G}}_{i} \Psi=0,\, \, \widehat{\mathcal{H}}_{a} \Psi=0,
\end{equation}
is also a solution to this (Hamiltonian) constraint. For example, the Chern-Simons state is still a solution to these constrains but is no longer necessary verify that is also a solution to the Hamiltonian constraint. \\

Now, it suggest something analogous in the ADM formalism \cite{rosastes, peldan1}, see \cite{rosas2}. Thus, the cosmological function for the ADM formalism extended to triads is given by 
\begin{equation} \label{k18}
\Lambda = \frac{p^{ij} p_{ij}  - \frac{1}{2}  \left( p^{i}_{i} \right)^{2} - 4 \, e^{2}{}^{(3)}R }{8 \, e^{2}}.
\end{equation}
This cosmological function also solves the Hamiltonian constraint in this formalism and is a scalar function again.

\section{Relation between cosmological function}
Now, the core of this work is to show that the cosmological function above are related through the spacial Ricci scalar and this is what we are going to do in this section.\\

Indeed, the first thing that we have to do is to rewrite the Hamiltonian constraint of the Ashtekar's formalism. We know that the ADM formalism in triads and the Ashtekar's formalism are related trough a canonical transformation \footnote{This canonical transformation was given by Ashtekar himself around the year 1986.} \cite{ash1, ash2, ash3} given as 
\begin{equation} \label{i1}
K_{ai}:= i (A_{ai} - \Gamma_{ai}),
\end{equation}
where $K_{ai}$ is an auxiliary field related to the extrinsic curvature, $A_{ai}$ is the Asthekar's connection and $\Gamma_{ai}$ is the unique free-torsion spin-connection defined to annihilate the field $\tilde{E}^{ai}$:
\begin{equation} \label{i2}
\mathcal{D}_{a}\tilde{E}^{bi}:= \partial_{a} \tilde{E}^{ai} + \Gamma_{ac}^{b} \tilde{E}^{ci} - \Gamma_{ac}^{c} \tilde{E}^{bi} + \epsilon^{ijk} \Gamma_{aj} \tilde{E}_{i}^{b}=0.
\end{equation}
Note that 
\begin{equation} \label{i3}
\mathcal{D}_{a}\tilde{E}^{ai}= \partial_{a} \tilde{E}^{ai} + \epsilon^{ijk} \Gamma_{aj} \tilde{E}_{k}^{a}=0
\end{equation}
Now, the curvature associate to $A_{ai}$ can be written as 
\begin{align*}
F_{ab \, i}&= 2 \partial_{[a}(\Gamma_{b]\, i} - i K_{b]\,i} ) + \epsilon_{ijk}\left(\Gamma_{a}^{j} \Gamma_{b}^{k} - i \Gamma_{a}^{j}K_{b}^{k} -i K_{a}^{j} \Gamma_{b}^{k} - K_{a}^{j} K_{b}^{k} \right)\\
&= 2 \partial_{[a} \Gamma_{b]i} +  \epsilon_{ijk} \Gamma_{a}^{j} \Gamma_{b}^{k} - 2i \partial_{[a}K_{b]i} - \epsilon_{ijk} K_{a}^{j} K_{b}^{k} - i \epsilon_{ijk} \left(\Gamma_{a}^{j}K_{b}^{k} - \Gamma_{b}^{j}K_{a}^{k} \right)\\
&= f_{abi} - 2i \left( \partial_{[a}K_{b]i} + \epsilon_{ijk} \Gamma_{[a}^{j}K_{b]}^{k} \right) - \epsilon_{ijk} K_{a}^{j} K_{b}^{k}.
\end{align*}
Thus
\begin{equation} \label{i5}
 F_{ab}^{i}=f_{ab}^{i} - 2i \mathcal{D}_{[a} K_{b]}^{i} - \epsilon^{ijk} K_{aj} K_{b k},
\end{equation}
where $f_{abi}$ is the curvature associate to $\Gamma_{ai}$ defined by: $f_{abi}:= 2 \partial_{[a} \Gamma_{b]i} +  \epsilon_{ijk} \Gamma_{a}^{j} \Gamma_{b}^{k}$.
It follows that the Hamiltonian constraint Eq. \eqref{k4} can be rewritten as 
\begin{align*}
\mathcal{H}&= \frac{1}{2} \epsilon^{ijk} \tilde{E}_{i}^{a} \tilde{E}_{j}^{b} (f_{ab k} - 2i \mathcal{D}_{[a} K_{b]\, k} - \epsilon_{klm} K_{a}^{l} K_{b}^{m})- \varphi\, \text{det} \tilde{E}^{ai}\\
&=  \frac{1}{2} \epsilon^{ijk} \tilde{E}_{i}^{a} \tilde{E}_{j}^{b} f_{ab k} - i \epsilon^{ijk} \tilde{E}_{i}^{a} \tilde{E}_{j}^{b} \mathcal{D}_{[a} K_{b]\, k} -  \tilde{E}_{i}^{a} \tilde{E}_{j}^{b}  K_{a}^{[i} K_{b}^{j]} - \varphi \, \text{det} \tilde{E}^{ai} \approx 0 .
\end{align*}
Ignoring surface terms we have 
\begin{equation} \label{i7}
\mathcal{H} =  \frac{1}{2} \epsilon^{ijk} \tilde{E}_{i}^{a} \tilde{E}_{j}^{b} f_{ab k}-  \tilde{E}_{i}^{a} \tilde{E}_{j}^{b}  K_{a}^{[i} K_{b}^{j]} - \varphi \, \text{det} \tilde{E}^{ai} \approx 0 
\end{equation}
Something analogous was done  by P. Peldan in \cite{peldan1}.
Now we define a version undensized of the variable $\tilde{E}^{ai}$ and its inverse, i.e., 
\begin{equation} \label{i8}
\tilde{E}^{ai}=: e e^{ai} \quad e_{bi}e^{ai}= \delta_{b}^{a} \quad e:= \text{det} e_{ai}= \sqrt{\text{det} \tilde{E}^{ai}}. 
\end{equation}
With this relation in hand we can write the scalar constraint Eq.\eqref{i7} in terms of triads $e^{ai}$ and its determinant. It follows that
$$\mathcal{H} =  \frac{1}{2} \epsilon^{ijk} e^{2} e_{i}^{a} e_{j}^{b} f_{ab k}-  e^{2} e_{i}^{a} e_{j}^{b} K_{a}^{[i} K_{b}^{j]} - \varphi e^{2} \approx 0. $$
Thus, the cosmological function for the Ashtekar's formalis can now be written as 
\begin{equation} \label{i17}
\varphi:=  \frac{1}{2} \epsilon^{ijk}  e_{i}^{a} e_{j}^{b} f_{ab k}-   e_{i}^{a} e_{j}^{b} K_{a}^{[i} K_{b}^{j]} .
\end{equation}
Note that
\begin{equation} \label{i18}
\epsilon^{ijk}  e_{i}^{a} e_{j}^{b} f_{ab k}= e_{i}^{a} e_{j}^{b} \Omega_{ab}^{ij},
\end{equation}
where $\Omega_{ab}^{ij}=-\Omega_{ab}^{ji}$ is a Lorentz connection. It is well know that the spacial Ricci scalar can be related to this connection by the following equation 
\begin{equation} \label{i19}
^{(3)\, } R:= e_{i}^{a} e_{j}^{b} \Omega_{ab}^{ij}.
\end{equation}
With this the cosmological function Eq. \eqref{i17} can be rewritten as
\begin{equation} \label{i17}
\varphi:=  \frac{1}{2} {} ^{(3)\, } R -   e_{i}^{a} e_{j}^{b} K_{a}^{[i} K_{b}^{j]} .
\end{equation}
This is a more familiar equation with relation to the cosmological function for the ADM formalism, see Eq. \eqref{k18}. Indeed, we can always solve for the curvature in Eq. \eqref{k18} to get 
\begin{equation} \label{i21}
 ^{(3)} R= - 2 \Lambda + \frac{1}{4 \, e^{2}} \left(  p^{ij} p_{ij}  - \frac{1}{2}  (p^{i}_{i})^{2} \right).
\end{equation}
From Eqs. \eqref{i21} and \eqref{i17} we obtain 
$$\varphi =  \frac{1 }{2} \left[  - 2 \Lambda + \frac{1}{4 \, e^{2}} \left(  p^{ij} p_{ij}  - \frac{1}{2}  (p^{i}_{i})^{2}  \right)\right]-e_{i}^{a} e_{j}^{b} K_{a}^{[i} K_{b}^{j]}. $$
Thus
\begin{equation} \label{i22}
\varphi = - \Lambda + \frac{1}{8 \, e^{2}}\left(  p^{ij} p_{ij}  - \frac{1}{2}  (p^{i}_{i})^{2}  \right) -  e_{i}^{a} e_{j}^{b} K_{a}^{[i} K_{b}^{j]}.
\end{equation}
The above equation represents the first relation between the two proposed cosmological functions. What follows is to write this equation more compactly.
Now, note that we can define the field $K_{ai}$ as \cite{peldan1, nicoli}
\begin{equation} \label{i23}
K_{a}^{i} = \frac{1}{ 2 \, e } \left(p_{a}^{i} - \frac{1}{2} (p_{k}^{k}) e_{a}^{i}\right).
\end{equation}
We can always define $p^{ij}= p^{ai}e_{a}^{j}$, then, 
\begin{equation} \label{i23a}
p^{ij}p_{ij}= p^{ai}e_{a}^{j} p_{bi}e_{j}^{b}= p^{ai} p_{bi} \delta_{a}^{b} = p^{ai}p_{ai}, \quad p:= p^{ai}e_{ai}= p^{ij}e_{j}^{a} e_{ai} = p_{i}^{i}.
\end{equation}
From Eqs. \eqref{i23} and \eqref{i23a} is straightforward to obtain an explicit expression for $ e_{i}^{a} e_{j}^{b}K_{a}^{[i} K_{b}^{j]}$. Thus, on one hand we have
$$e_{i}^{a} e_{j}^{b} K_{a}^{i} K_{b}^{j}= e_{i}^{a} e_{j}^{b} \frac{1}{4 e^{2}}\left(p_{a}^{i}-  \frac{1}{2} p e_{a}^{i}\right)\left(p_{b}^{j}-  \frac{1}{2} p e_{b}^{j}\right) = \frac{1}{4 e^{2}} \left[ e_{i}^{a} e_{j}^{b} p_{a}^{i}p_{b}^{j}- \frac{p}{2}(e_{b}^{j}p_{a}^{i} e_{i}^{a} e_{j}^{b} +e_{a}^{i} p_{b}^{j} e_{i}^{a} e_{j}^{b}) \right.$$
$$\left. + \frac{1}{4} p^{2} e_{i}^{a} e_{j}^{b} e_{a}^{i} e_{b}^{j} \right] = \frac{1}{4 e^{2}} \left[ p_{i}^{i} p_{j}^{j}- \frac{p}{2} (p_{i}^{i} \delta_{j}^{j} + p_{j}^{j} \delta_{i}^{i}) + \frac{1}{4} p^{2} \delta_{i}^{i} \delta_{j}^{j} \right]= \frac{1}{4 e^{2}} \left( p^{2} - 3 p^{2} + \frac{9}{4} p^{2} \right). $$
So
\begin{equation}\label{i24}
 e_{i}^{a} e_{j}^{b} K_{a}^{i} K_{b}^{j}= \frac{1}{16 \, e^{2}} \, p^{2}.
\end{equation}
On the other hand we have
\begin{align*}
e_{i}^{a} e_{j}^{b} K_{a}^{j} K_{b}^{i} &= \frac{1}{4 e^{2}} \left[ e_{i}^{a} e_{j}^{b} p_{a}^{j} p_{b}^{i} - \frac{p}{2} e_{i}^{a} e_{j}^{b} ( p_{a}^{j} e_{b}^{i}  + p_{b}^{i} e_{a}^{j}) + \frac{1}{4} p^{2} e_{i}^{a} e_{j}^{b} e_{a}^{j} e_{b}^{i}\right] \\
& = \frac{1}{4 e^{2}} \left[ p_{i}^{j} p_{j}^{i}  - \frac{p}{2} (p_{i}^{j} \delta_{j}^{i} + p_{j}^{i} \delta_{i}^{j}) + \frac{1}{4} p^{2} \delta^{ii} \right]=  \frac{1}{4 e^{2}} \left( p^{ij} p_{ij}  - p^{2} + \frac{3}{4} p^{2} \right) .
\end{align*}
So
\begin{equation} \label{i25}
 e_{i}^{a} e_{j}^{b} K_{a}^{j} K_{b}^{i} =   \frac{1}{4 e^{2}} \left( p^{ij} p_{ij}  - \frac{1}{4} p^{2} \right).
\end{equation}
It follows that 
$$e_{i}^{a} e_{j}^{b}K_{a}^{[i} K_{b}^{j]}  = \frac{1}{2} \left[ e_{i}^{a} e_{j}^{b} K_{a}^{i} K_{b}^{j} - e_{i}^{a} e_{j}^{b} K_{a}^{j} K_{b}^{i} \right] = \frac{1}{2}  \left[  \frac{1}{16 \, e^{2}} \, p^{2}-  \frac{1}{4 e^{2}} \left(p^{ij} p_{ij}  - \frac{1}{4} p^{2}\right) \right].$$
Therefore
\begin{equation} \label{i26}
 e_{i}^{a} e_{j}^{b}K_{a}^{[i} K_{b}^{j]}  = - \frac{1}{8 e^{2}} \left(p^{ij}p_{ij}- \frac{1}{2} (p_{k}^{k})^{2} \right).
\end{equation}
As an consequence we have 
$$\varphi = - \Lambda + \frac{1}{8 \, e^{2}} \left(  p^{ij} p_{ij}  - \frac{1}{2}  (p^{i}_{i})^{2} \right) +  \frac{1}{8 e^{2}} \left(p^{ij}p_{ij}- \frac{1}{2} (p_{i}^{i})^{2} \right).$$
Hence the cosmological function in the Ashteka's formalism can be written as 
\begin{equation} \label{i27}
 \varphi = - \Lambda + \frac{1}{4 \, e^{2}} \left(  p^{ij} p_{ij}  - \frac{1}{2}  (p^{i}_{i})^{2}  \right).
\end{equation}
This is a more compact relation between the proposed cosmological functions. \\
But, note that from Eq. \eqref{k18} we have that 
\begin{equation}
 p^{ij} p_{ij}  - \frac{1}{2}  (p^{i}_{i})^{2}= 8 e^{2} \Lambda + 4 e^{2} \, {}^{(3)}R.
\end{equation}
Then, we can write 
$$\varphi = - \Lambda + \frac{1}{4 \, e^{2}} \left( 8 e^{2} \Lambda + 4 e^{2} \, {}^{(3)}R \right)= - \Lambda + 2 \Lambda + {}^{(3)}R.$$
Thus we have a more compact relation
\begin{equation} \label{ilol}
\varphi = \Lambda + {}^{(3)}R,
\end{equation}
Indeed, the above equation represents the most compact relation between the prosed cosmological functions that we found.\\
Note that the cosmological functions in both formalism are the same up to the spacial Ricci scalar.

We also found a way to relate this cosmological functions through the extrinsic curvature. To do this, we begin from the fact that the equation Eq. \eqref{i27} can be written as 
\begin{equation} \label{i27a}
\varphi = - \Lambda + \frac{1}{4 \, e^{2}} \left(  p^{ai} p_{ai}  - \frac{1}{2}  p^{2} \right). 
\end{equation}
Then, from Eq. \eqref{i23} we have $p^{ai}= 2 K^{ai} e + \frac{1}{2}\, p e^{ai}$, so 
\begin{equation} \label{i28}
 p^{ai} p_{ai} = 4 e^{2} K_{ai}K^{ai} + 2 p \, e K^{ai}e_{ai} + \frac{p^{2}}{4} \delta_{i}^{i}= 4 K_{ai}K^{ai} + 2 p \, e K + \frac{3}{4}p^{2},  
\end{equation}
where we have defined $K:= K^{ai}e_{ai}$.\\
Now an alternative definition to $K_{a}^{i}$ is \cite{capo2}:
\begin{equation} \label{i29}
K_{a}^{i}:= K_{ab} e^{bi} + J_{ab}e^{bi},
\end{equation}
where $K_{ab}=K_{(ab)}$ is the extrinsic curvature and  $J_{ab}=J_{[ab]}$. Because of the ADM formalism extended to triads the annihilation of $J_{ab}$ is equivalent to that the internal constraint is  satisfied [see Eq. \eqref{cp1}]. Hence, 
$$K_{ai}K^{ai}= K_{ab} K^{ac} \delta_{c}^{b} + 2 e^{-1} K_{ab} J^{ac} \delta_{c}^{b} + e^{-2} J_{ab} J^{ac} \delta_{c}^{b}.$$
So
\begin{equation} \label{i30}
K_{ai}K^{ai}=K_{ab}K^{ab} + 2 e^{-1} K^{ab} J_{ab} + e^{-2} J_{ab} J^{ab}.
\end{equation}
It follows that 
$$p^{ai} p_{ai} = 4  e^{2} ( K_{ab}K^{ab} + 2 e^{-1} K^{ab} J_{ab} + e^{-2} J_{ab} J^{ab})  + 2 p \, e K + \frac{3}{4}p^{2}.$$
Next, note that 
$$p:= e_{ai}p^{ai}= 2 K^{ai} e_{ai} e + \frac{p}{2} e_{ai} e^{ai}.$$
Hence 
$$ K=- \frac{1}{4 \, e} p \Rightarrow K^{2}= \frac{1}{16 e^{2}} p^{2}.$$
Then  
$$p^{ai} p_{ai} =4  e^{2} \left(K_{ab}K^{ab} + 2 e^{-1} K^{ab} J_{ab} + e^{-2} J_{ab} J^{ab} \right) + 2\, e\,  (- 4 e K) K + \frac{3}{4} (16 e^{2} K^{2}).$$
So
\begin{equation} \label{i31}
p^{ai} p_{ai} =4  e^{2} \left( K_{ab}K^{ab} + 2 e^{-1} K^{ab} J_{ab} + e^{-2} J_{ab} J^{ab} \right) + 4e^{2}K^{2}.
\end{equation}
Thus
$$\varphi = - \Lambda + \frac{1}{4 \, e^{2}} \left[ 4  e^{2} \left( K_{ab}K^{ab} + 2 e^{-1} K^{ab} J_{ab} + e^{-2} J_{ab} J^{ab} \right)  + 4e^{2}K^{2} - \frac{1}{2}  (16 e^{2} K^{2}) \right], $$
and 
\begin{equation} \label{i32}
 \varphi = - \Lambda + \frac{1}{4 \, e^{2}} \left[ 4   e^{2} \left( K_{ab}K^{ab} + 2 e^{-1} K^{ab} J_{ab} + e^{-2} J_{ab} J^{ab}\right)  - 4e^{2}K^{2}  \right].
\end{equation}
Therefore 
\begin{equation} \label{i33}
\varphi \approx  - \Lambda  + K_{ab}K^{ab}-K^{2} ,
\end{equation}
where $\approx $ denote modulo the internal constriction. But when we do an restriction to $X_{phy.}$ we obtain that
\begin{equation}
K_{ab}K^{ab} - K^{2}\approx {}^{(3)} R.
\end{equation}
In conclusion
\begin{equation} \label{i34}
\varphi \approx  - \Lambda  + {}^{(3)}R. 
\end{equation}
This represents the most compact relation between the proposed cosmological functions when we restrict our attention to the physical phase space $X_{phy}$.\\

\section{Concluding Remarks}
Firs of all we need to say that the relations given by Eqs. \eqref{ilol} and \eqref{i34} do not have to be equal, we find Eq. \eqref{ilol} considering the whole phase space, i.e. $Met(\Sigma)$, while one finds Eq. \eqref{i34} just when we restrict our attention to the physical phase space $X_{phy.}$. Indeed, what led us to Eq. \eqref{i34} was that we were looking for a more geometrical form of relate the proposed cosmological functions. \\

Although the ansatz $\Lambda$ and $\varphi$, Eqs. \eqref{k18} and \eqref{k14} respectively, are obtained directly by one simple computation, the interpretation for these cosmological functions is not clear for us. We have to admit that at this moment is not clear why the cosmological function are related through the three-dimensional Ricci scalar. Perhaps a more geometric interpretation can be given from this relation in terms of extrinsic curvature, it is well know that the extrinsic curvature only makes sense when we have a manifold embedded on another of higher dimension and this is what we have in GR, see Eq. \eqref{i33}.\\

We hope to have a more precise interpretation of this cosmological function and their relation in the coming years. 
%And we really hope that this work have something to do with the cosmological constant problem that is one of the special motivation to the developed of the Krasnov's theories. 


\begin{thebibliography}{99}
\bibitem{wald} R. M. Wald, \textit{General Relativity}, The University of Chicago Press, Chicago (1984).
\bibitem{baez} J. Baez and J. P. Muniainn, \textit{Gauge Fields, Knots and Gravity}, World Scientific, Singapore (1994).
\bibitem{gambi} R. Gambini and J. Pullin, \textit{Loops, Knots, Gauge Theories and Quantum Gravity}, Cambridge University Press, Cambridge, UK (2000).
\bibitem{dicke} R. H. Dicke, ``Republication of The Theoretical significance of experimental relativity", \textit{Gen. Rel. and Grav.}, \textbf{51}, 1-31, (2019).
\bibitem{mann} P. D. Mannheim, ``Linear potentials an galactic rotation curves", \textit{Ast. J.}, \textbf{419}, 150 (1993).
\bibitem{kr3} K. Krasnov, ``Non-metric Gravity: I. Field Equations", \textit{Class. Quantum Grav}. \textbf{25} 025001, (2008)
\bibitem{kr2} K. Krasnov, ``On Deformations of Ashtekar's Constraint Algebra", \textit{Phys. Rev. Lett.} \textbf{100}, 081102.  (2007).
\bibitem{rosas1} R. Rosas-Rodríguez, ``Hamiltonian Constrain of Gravity with Cosmological Function", \textit{AIP. Conf. Proc.} \textbf{1548}, 191 (2013).
\bibitem{rosas2} R. Rosas-Rodríguez, ``Cosmological Functions in ADM and Ashtekar's Representations of Gravity", en preparación.
\bibitem{pleva} J. F. Plebanski, ``On the Separation of Einsteinian Substructures", \textit{J. Math. Phys.} \textbf{18}, 2511 (1977).
\bibitem{kr1} K. Krasnov, ``Renormalizable Non-Metric Quantum Gravity?", arXiv:hep-th/0611182 (2007).
\bibitem{capo} R. Capovilla, J. Dell, T. Jacobson and L. Mason, ``Selfdual 2-forms and Gravity", \textit{Class. Quantum Grav.} \textbf{8}, 41 (1990).
\bibitem{rosas3} R. Rosas-Rodríguez, ``Note on the Constraint Algebra of Modified Gravity Theories", \textit{AIP. Conf. Proc.} \textbf{1287}, 85 (2010).
\bibitem{capomon} R. Capovilla, M. Montesinos and M. Velázquez, ``Minimally Modified Self-dual 2-forms Gravity", \textit{Class. Quantum Grav}, \textbf{27}, 145011 (2010).
\bibitem{rosastes} R. Rosas-Rodríguez, \textit{Estructuras Hamiltonianas para Campos Clásicos}, Tesis de Doctorado BUAP, Puebla (2007).
%\bibitem{rosas4} R. Rosas-Rodríguez, ``On the Chern-Simons State in General Relativity and Modified Gravity Theories", \textit{J. Phys.: Conf. Ser.} \textbf{545} 012013 (2014).
\bibitem{peldan1} P. Peldan, ``Actions for Gravity, with Generalizations: A Review", \textit{Class. Quantum Grav}. \textbf{11} 1087, (1994).
\bibitem{ash1} A. Ashtekar, ``New Variables for Classical and Quantum Gravity", \textit{Phys. Rev. Lett. }\textbf{57}, 2244 (1986).
\bibitem{ash2} A. Ashtekar, ``New Hamiltonian Formulation of General Relativity", \textit{Phys. Rev.} D \textbf{36}, 1587 (1987).
\bibitem{ash3} A. Ashtekar, (notes prepared in collaboration with R.S. Tate), \textit{Lectures on Non-Perturbative Canonical Gravity}, World Scientific, Singapore, (1991).
%\bibitem{rove2} C. Rovelli, \textit{Quantum Gravity}, Cambridge University Press, Cambridge, UK (2004).
\bibitem{nicoli} H. Nicolai, H. J. Matschull, ``Aspects of Canonical Gravity and Supergravity", \textit{Journal of Geometry and Physics}. \textbf{11}.1-4: 15-62 (1993).
\bibitem{capo2}  R. Capovilla, ``Canonical Gravity", Proc. IV Mexican Workshop on Particles and Fields, p 217, Mérida (1993).
\bibitem{kr4} K. Krasnov, Y. Shtanov, ``Non-metric Gravity: II. Spherically Gymmetric Golution, Missing Mass and Redshifts of Quasars", \textit{Class. and Quantum Grav.}, \textbf{25}, 025002 (2007).
\end{thebibliography}
\end{document}